\documentclass{agujournal}
\journalname{JGR-Atmospheres}
\usepackage{gensymb}
\usepackage{lineno}
\usepackage{amsmath}
\usepackage{wasysym}
\usepackage{float}
\usepackage{multirow}
 \usepackage{graphicx}
  \setkeys{Gin}{draft=false}
\begin{document}

%
%

\title{The Snowball Stratosphere}
%
%
\authors{R.J. Graham{\affil{1}}, Tiffany A. Shaw{\affil{2}}, Dorian S. Abbot{\affil{2}}}
\affiliation{1}{Department of Physics,
  The University of Oxford, Oxford, UK}
\affiliation{2}{Department of Geophysical Sciences, University of
  Chicago, Chicago, IL, USA}
\begin{keypoints}
\item The simulated snowball stratosphere with or without ozone displays weaker circulation than the modern case.
\item The snowball stratosphere with ozone has weaker wave forcing and a stronger jet than the modern and has no sudden stratospheric warmings.
\item Changes to stratospheric circulation during the snowball do not impact pCO$_2$ estimated from proxy data unless pCO$_2$ was much greater than pO$_2$.
\end{keypoints}
%
%

\begin{abstract}
According to the Snowball Earth hypothesis, Earth has experienced periods of low-latitude glaciation in its deep past. Prior studies have used general circulation models (GCMs) to examine the effects such an extreme climate state might have on the structure and dynamics of Earth's troposphere, but the behavior of the stratosphere has not been studied in detail. Understanding the snowball stratosphere is important for developing an accurate account of the Earth's radiative and chemical properties during these episodes. Here we conduct the first analysis of the stratospheric circulation of the Snowball Earth using ECHAM6 general circulation model simulations. In order to understand the factors contributing to the stratospheric circulation, we extend the Statistical Transformed Eulerian Mean framework. We find that the stratosphere during a snowball with prescribed modern ozone levels exhibits a weaker meridional overturning circulation, reduced wave activity, stronger zonal jets, and is extremely cold relative to modern conditions. Notably, the snowball stratosphere displays no sudden stratospheric warmings. Without ozone, the stratosphere displays slightly weaker circulation, a complete lack of polar vortex, and even colder temperatures. We also explicitly quantify for the first time the cross-tropopause mass exchange rate and stratospheric mixing efficiency during the snowball and show that our values do not change the constraints on CO$_2$ inferred from geochemical proxies during the Marinoan glaciation ($\sim$635 Ma), unless the O$_2$ concentration during the snowball was orders of magnitude less than the CO$_2$ concentration.
\end{abstract}



%
%

%

%
%
\section{Introduction}
\label{intro}

The snowball Earth hypothesis proposes that during the Neoproterozoic glaciations ($\sim$710 and $\sim$635 Ma) ice covered most or all of Earth's oceans and continents from the poles to very low latitudes, possibly to the equator \citep{Kirschvink92,Hoffman98}. During this time, silicate weathering would be greatly reduced, allowing CO$_2$ from volcanic outgassing to build up to the extremely high levels needed to melt the tropical ice and reverse the global glaciation \citep{Walker-Hays-Kasting-1981:negative, Kirschvink92, Hoffman98}. Although there are alternative models that attempt to explain observations of low-latitude Neoproterozoic glaciations without invoking a full snowball \citep{Hyde00,Chandler:2000,Peltier07,allen2008,Micheels08,Liu:2010p1977,Abbot-et-al-2011:Jormungand,yang2011a,yang2011b,rose2015stable}, this study focuses on the asymptotic case of a full, global glaciation.

How life survived the snowball events and how deglaciation occurred are important questions that haven't been fully resolved \citep{Pierrehumbert-et-al-2010:neoprot,hoffman2017snowball}. Atmospheric circulation during a snowball influences both of these issues through heat transport; the lofting of dust, sometimes referred to as cryoconite, which affects surface albedo and the formation of holes in the ice; and clouds, which can generate greenhouse warming. So far most investigations have focused on the snowball troposphere. For example, previous work has shown that 1.) the tropopause in a snowball is lower \citep{Pierrehumbert04,Pierrehumbert05,abbot2014}; 2.) the Hadley circulation is more intense \citep{Pierrehumbert04,Pierrehumbert05,Voigt2012,voigt2013dynamics,abbot13-snowball-circulation}, thermally indirect in the annual mean \citep{Pierrehumbert04,Pierrehumbert05,Abbot-Pierrehumbert-2009:mudball}, and strongly influenced by vertical diffusion of zonal momentum \citep{Voigt2012,voigt2013dynamics}; and 3.) baroclinic eddy dry static energy transport is proportionally more important than in its modern counterpart \citep{Pierrehumbert04,Pierrehumbert05}. The only investigation of the snowball's stratosphere that we have encountered is that of \citet{yang2012radiative}, which examined the radiative effect of ozone on the snowball atmosphere's temperatures and found that reduced ozone leads to a cooler surface. The stratospheric circulation during a snowball, however, has not been investigated in detail. 

The dynamics of the snowball stratosphere and stratosphere-troposphere exchange have important consequences for the interpretation of a widely cited piece of geochemical evidence for the snowball. The atmospheric $\Delta^{17}$O ($= \delta^{17}$O$-0.52\times \delta^{18}$O), a measure of the mass-independent fractionation of oxygen isotopes 17 and 18, reflects changes in the concentration of CO$_2$ via a stratospheric photochemical reaction network linking O$_3$, O$_2$, and CO$_2$ \citep{yung1991isotopic,yung1997carbon}. With other quantities held constant, greater CO$_2$ content leads to more negative $\Delta^{17}$O. There are large negative anomalies in $\Delta^{17}$O fractionation recorded in sulfates that were produced during and immediately after the second proposed Neoproterozoic glaciation (the Marinoan glaciation, 635 Ma), which have been interpreted as evidence for a very high CO$_2$ partial pressure from volcanism in the absence of silicate weathering \citep{Bao08, Bao09,Cao2013}. The inferred quantity of CO$_2$ depends on the circulation within the stratosphere and the stratosphere-troposphere exchange because the signal observed in rocks derives from fractionation in the upper stratosphere that dilutes as it propagates into the troposphere. 

The modern stratospheric circulation is {predominantly} wave-driven and strongly impacts stratospheric temperatures and ozone abundance {\citep{andrews1987middle,Haynes2005stratospheric,maycock2013circulation,butchart2014brewer}}. In particular, waves generated in the troposphere interact with the {winter hemisphere's} stratospheric polar vortex and converge in the stratosphere driving the Brewer-Dobson circulation (poleward motion in midlatitudes and descent over the pole) via a Coriolis torque. When the wave-driving is particularly strong it can reverse the sign of the stratospheric polar vortex from eastward to westward, an event called a sudden stratospheric warming \citep{butler2015defining}. Future projections suggest an acceleration of the stratospheric circulation under warming induced by greenhouse gas forcing \citep{mclandress2009simulated} due to enhanced wave breaking in the lower stratosphere. How the different factors affecting the stratospheric circulation, e.g. wave driving, are impacted during cold conditions like the snowball has not been investigated.

Here, we quantify the stratospheric circulation during a snowball using a general circulation model (GCM). {With modern ozone, the simulated snowball stratosphere displays a weaker circulation, weaker wave activity, stronger jets, an absence of sudden stratospheric warmings, and colder temperatures compared to conditions in our modern simulation. In the absence of ozone, the stratospheric temperatures are colder, circulation is weaker, wave-driving is weaker, and stratospheric zonal jets are non-existent.} We also find that these changes to the circulation probably do not change geochemical proxy-based estimates of CO$_2$ during the snowball that were calculated with the implicit assumption that the snowball stratosphere behaved like the modern, unless the snowball's O$_2$ concentration was much lower than its CO$_2$ concentration. In the remainder of the paper, we describe the simulations and methods we used to diagnose and analyze the properties of the snowball stratosphere (Section \ref{methods}), summarize our results (Section \ref{results}), and speculate about implications of this work while also describing future directions (Section \ref{discussion}).

\section{Methods}
\label{methods}
\subsection{Model Configuration}
We ran the ECHAM6 \citep{giorgetta2013atmospheric} GCM with boundary conditions set to represent the pre-industrial modern Earth and several variants of an idealized hard snowball Earth. {We ran each simulation for a five year spin-up period, checked for equilibrium to within 1 W m$^{-2}$ at the top of the atmosphere, and ran the simulations for an additional five years to generate the data used in this study.} All simulations have a 360-day year, zero eccentricity, and obliquity equal to 23.4$^{\circ}$, and they use a modern continental configuration with a 50m deep slab ocean. {We do not include ocean energy transport because it does not affect the snowball circulation, and our simulated modern stratospheric circulation agrees well with ERA reanalysis without it.} The slab ocean is covered in sea ice in the snowball simulations. We use a prescribed seasonally-varying modern ozone profile, except in the simulation that lacks ozone. We consider a simulation without ozone because O$_2$ levels are poorly constrained during the Neoproterozoic and may have been one or two orders of magnitude lower than today \citep{canfield2005early,yang2012radiative,laakso2017theory}. The snowball simulations have CO$_2$ concentrations of 280 ppmv and $10^4$ ppmv in order to test the effect of CO$_2$ build up over time during the snowball due to weakened silicate weathering. We also cover the land surface with glaciers in the snowball simulations and use a constant albedo of 0.60 for both glaciers and sea ice. {The glaciers do not change the topography of the land.} We ran one of the snowball simulations with flattened topography to test the influence of topography on circulation and eddy activity. A summary of the different simulations can be found in Table \ref{tab:sims}.

\subsection{Stratospheric circulation}

The stratospheric circulation is typically quantified using the transformed Eulerian mean (TEM) circulation, which accounts for the impact of wave driving on the circulation \citep{andrews1987middle}. However, the TEM is only approximate, as it assumes small amplitude eddies. A more accurate representation of the stratospheric circulation is the dry isentropic circulation, which does not require any eddy amplitude approximation (see e.g. \citet{pauluis2008global}). \citet{pauluis2011statistical} introduced the Statistical Transformed Eulerian Mean (STEM) as a simple and accurate approximation of the dry isentropic circulation that allows one to separate the mean and eddy components. The main assumption for the STEM framework is that the eddy variance is finite and Gaussian. In the limit of small eddy variance STEM converges to the TEM formulation.

Like the TEM circulation, the STEM circulation can be decomposed into mean and eddy components:
\begin{linenomath}
\begin{align}
\begin{split}
\Psi_{\text{STEM}}(\theta,\phi)&= \Psi_{\text{eddy}}(\theta,\phi)+\Psi_{\text{mean}}(\theta,\phi)\\
&=\int_{-\infty}^\theta \int_0^\infty \frac{2\pi a \cos\phi}{g}\frac{\overline{v'\theta'}(\theta-\overline{\theta})}{\sqrt{2\pi\overline{\theta'^2}^3}}\exp\bigg[\frac{-(\theta-\overline{\theta})^2}{2\overline{\theta'^2}}\bigg]d\Tilde{p}d\Tilde{\theta}\\
&+\int_{-\infty}^\theta \int_0^\infty \frac{2\pi a \cos\phi}{g}\frac{\overline{v}}{\sqrt{2\pi\overline{\theta'^2}}}\exp\bigg[\frac{-(\theta-\overline{\theta})^2}{2\overline{\theta'^2}}\bigg]d\Tilde{p}d\Tilde{\theta} 
\end{split}\end{align}\end{linenomath}
where $\phi$ is latitude, $v$ is the meridional wind velocity, $\theta$ is potential temperature, g is surface gravity, p is pressure, overbars represent zonal and monthly averages, and primes represent deviations from those averages. $\Psi_{\text{mean}}$ approximates the Eulerian mean circulation, including the three-cell structure in isentropic coordinates (Hadley, Ferrel, and polar cells, see Figures \ref{fig:era} and \ref{fig:modern}, although the polar cells are unresolved at that contour interval). $\Psi_{\text{eddy}}$ approximates the circulation driven by eddy heat fluxes, which is analogous to a Stokes drift in isentropic coordinates.

As mentioned previously, diffusion has been found to dominate the tropospheric circulation in simulations of the snowball \citep{Voigt2012,voigt2013dynamics}. The mean circulation $\Psi_\text{mean}$ depends on the mean meridional flow $\overline{v}$. Here we extend the STEM formulation to specifically account for advection of mean meridional momentum, eddy momentum flux convergence, and unresolved forces (which includes diffusion). Following \citet{voigt2013dynamics} we can use the zonal-momentum budget in the free atmosphere to decompose the mean meridional flow into components ($\overline{v} = \overline{v}_\text{adv} + \overline{v}_\text{EMFC} + \overline{v}_\text{{F}}$) related to mean advection ($\overline{v}_\text{adv}$), eddy momentum flux convergence (EMFC, $\overline{v}_\text{EMFC}$), and residual ($\overline{v}_\text{{F}}$):
\begin{linenomath}
\begin{align}
\frac{\partial \overline{u}}{\partial t} = 0 &= f\overline{v}+\overline{\zeta}\overline{v}-\overline{\omega}\frac{\partial \overline{u}}{\partial p} - \frac{1}{a\cos^2\phi}\frac{\partial(\overline{u'v'}\cos^2\phi)}{\partial\phi}-\frac{\partial\overline{u'\omega'}}{\partial p}+\overline{{F}}^u_D\\
\overline{v}_\text{adv} &= \frac{1}{f}\bigg(-\overline{\zeta}\overline{v}+\overline{\omega}\frac{\partial \overline{u}}{\partial p}\bigg)\\
\overline{v}_\text{EMFC} &=\frac{1}{f}\bigg(\frac{1}{a\cos^2\phi}\frac{\partial(\overline{u'v'}\cos^2\phi)}{\partial \phi}+\frac{\partial\overline{u'\omega'}}{\partial p}\bigg)\\
\overline{v}_\text{{F}} &= -\frac{\overline{{F}}_D^u}{f}
\end{align}\end{linenomath}
where $f = 2\Omega\sin{\phi}$, $\zeta$ is relative vorticity, $\omega$ is vertical pressure velocity, $p$ is pressure, $u$ is the zonal wind, and $\overline{F_D}^u$ is the inferred residual component of the zonal momentum budget, which we assume to be dominated by diffusion. With this decomposition, the STEM formulation for the mean circulation can be written as the sum of advection, EMFC, and residual contributions, $\Psi_\text{mean} = \Psi_{\text{adv}} + \Psi_{\text{EMFC}} + \Psi_{\text{{F}}}$, where:

\begin{linenomath}
\begin{align}
\Psi_{\text{adv}} &= \int_{-\infty}^\theta \int_0^\infty \frac{2\pi a \cos\phi}{g}\frac{\overline{v}_\text{adv}}{\sqrt{2\pi\overline{{\Tilde{\theta}}'^2}}}\exp\bigg[\frac{-({\Tilde{\theta}}-\overline{{\Tilde{\theta}}})^2}{2\overline{{\Tilde{\theta}}'^2}}\bigg]d\Tilde{p}d\Tilde{\theta}\label{eqn:psi_adv}\\
\Psi_{\text{EMFC}} &= \int_{-\infty}^\theta \int_0^\infty \frac{2\pi a \cos\phi}{g}\frac{\overline{v}_\text{EMFC}}{\sqrt{2\pi\overline{{\Tilde{\theta}}'^2}}}\exp\bigg[\frac{-({\Tilde{\theta}}-\overline{{\Tilde{\theta}}})^2}{2\overline{{\Tilde{\theta}}'^2}}\bigg]d\Tilde{p}d\Tilde{\theta}\label{eqn:psi_emfc}\\
\Psi_{\text{{F}}} &= \int_{-\infty}^\theta \int_0^\infty \frac{2\pi a \cos\phi}{g}\frac{\overline{v}_\text{{F}}}{\sqrt{2\pi\overline{{\Tilde{\theta}}'^2}}}\exp\bigg[\frac{-({\Tilde{\theta}}-\overline{{\Tilde{\theta}}})^2}{2\overline{{\Tilde{\theta}}'^2}}\bigg]d\Tilde{p}d\Tilde{\theta}\label{eqn:psi_diff}
\end{align}
\end{linenomath}

This decomposition therefore allows us to quantify the importance of diffusion for the stratospheric circulation. Note that because the meridional flow does not take into account the momentum balance at the surface these circulations do not close at the surface.

Finally, to better understand changes to the strengths of the ``shallow branches'' and ``deep branch'' of the eddy-driven circulation (e.g. \citet{plumb2002stratospheric}), following \citet{shaw2012tropical} we also decompose $\overline{v'\theta '}$ into contributions from planetary-scale (k=1-3) and synoptic-scale (k = 4-30) eddies and calculate the circulation for each. With these decompositions of the eddy-driven and mean components of $\Psi_\text{STEM}$, we can understand the drivers of the stratospheric circulation.
\subsection{The Eliassen-Palm flux}

In order to quantify the importance of wave driving in the stratosphere we follow \citet{kushner2004stratosphere} and calculate the Eliassen-Palm (EP) flux convergence in the vicinity of the winter stratospheric jet:
\begin{linenomath*}
\begin{align}
\begin{split}
    \int_{p_1}^{p_2}\int_{\phi_1}^{\phi_2}&\nabla\cdot\boldsymbol{F}\\
    &=a^{-1}\cos{\phi}\int_{p_1}^{p_2} F_{(\phi)} dp|_{\phi_2} - a^{-1}\cos{\phi}\int_{p_1}^{p_2} F_{(\phi)} dp|_{\phi_1}\\
    & + \int_{\phi_1}^{\phi_2}\cos{\phi} F_{(p)} d\phi|_{p_2} - \int_{\phi_1}^{\phi_2}\cos{\phi} F_{(p)} d\phi|_{p_1}
\end{split}
\end{align}\end{linenomath*}
where $F_{(\phi)} = a\cos{\phi} \Bigg[-\overline{u'v'}+(\frac{\partial\overline{u}}{\partial p})\overline{v'\theta '}/(\frac{\partial \overline{\theta}}{\partial p})\Bigg]$, $F_{(p)} = a\cos{\phi} \Bigg[(f+\overline{\zeta})\overline{v'\theta '}/(\frac{\partial \overline{\theta}}{\partial p})-\overline{u'\omega'}\Bigg]$, $\phi_1$ is $40.0^{\circ}$, $\phi_2$ is $90.0^{\circ}$, $p_1$ is the average pressure at the tropopause as defined by the 380 K isentrope in the modern case and the 285 K isentrope in the snowball cases, and $p_2$ is 1.0 hPa. With the integrated convergence of the EP flux, we can compare the propagation of wave activity between the troposphere and stratosphere in our modern and snowball simulations. This provides information useful for explaining differences in the structure and behavior of the zonal-mean flow of the stratosphere in our simulations.

\subsection{Tropospheric $\Delta^{17}$O}\label{sec:delta_17}
{$\Delta^{17}$O$_{O_2}$ is a measure of the degree of non-mass-dependent depletion of heavy oxygen ($^{17}$O and $^{18}$O) in O$_2$. This depletion occurs because the photochemical formation of ozone from O$_2$ in the upper stratosphere tends to produce O$_3$ molecules with mass independent enrichment of $^{17}$O and $^{18}$O \citep{yung1991isotopic,yung1997carbon}. This ozone fractionation signal is transferred to CO$_2$ through isotopic exchange reactions that move the excited species O($^1$D), which is released in UV photolysis of O$_3$ and therefore enriched in heavy oxygen, to CO$_2$ \citep{yung1991isotopic,yung1997carbon}. Transfer of heavy oxygen atoms from O$_2$ to O$_3$ to CO$_2$ leads to a depletion of heavy oxygen in O$_2$ (a negative $\Delta^{17}$O$_{O_2}$), and a larger reservoir of CO$_2$ allows for the storage of more heavy oxygens and correspondingly more negative $\Delta^{17}$O$_{O_2}$ \citep{yung1991isotopic,yung1997carbon}.} 

{The interpretation of anomalously negative $\Delta^{17}$O$_{O_2}$ in ancient minerals as evidence for high CO$_2$ during and immediately after the Neoproterozoic snowball requires that O$_2$ bearing this signal propagated down from the stratosphere into the troposphere without entirely losing its mass-independent fractionation.} In the rest of the study, we will use $\Delta^{17}$O to refer to the fractionation of tropospheric O$_2$ specifically unless otherwise noted. In \citet{Cao2013}, the heavy oxygen fractionation of tropospheric oxygen in steady state is given as:
\begin{linenomath}
\begin{align}
    \Delta^{17}\text{O} &= \delta^{17}\text{O}-0.52\times\delta^{18}\text{O}\\
    &= \frac{-\Phi(\rho)\gamma\Theta\tau}{1+\rho+\gamma\Theta\tau} \label{eq_delta_17}
\end{align}
\end{linenomath}
where $\rho$ is the ratio of pO$_2$/pCO$_2$ (pO$_2$ and pCO$_2$ are the concentrations of the respective gases), $\Phi(\rho)$ ($\equiv0.519\times(64+146\times(\rho/1.23))/(1+\rho/1.23)-7.1738$) is the difference of $\Delta^{17}$O between steady state O$_2$ and CO$_2$ in the photochemical reaction, $\tau$ is the residence time of oxygen ($\equiv \text{pO}_2/$F$_{op}$, and F$_{op}$ is the rate of oxygen production), $\gamma$ is the cross-tropopause mass exchange rate {(defined later in section \ref{sec:delta_17})}, and $\Theta$ is the stratospheric mixing efficiency {(defined later in section \ref{sec:delta_17})}. 

{Qualitatively, equation \ref{eq_delta_17} means that the degree to which the fractionation signal can propagate into the troposphere from the stratosphere is determined by 1.) the magnitude of fractionation in the steady-state reaction network in the upper stratosphere (a function of $\rho$); 2.) how well the depleted O$_2$ can mix from the upper stratosphere to the lower stratosphere (captured by $\Theta$); 3.) the rate of cross-tropopause mass exchange ($\gamma$) to carry the fractionation signal from the lower stratosphere into the troposphere; and 4.) the residence time of oxygen ($\tau$), since O$_2$'s residence time is determined by its rate of photosynthetic production, and photosynthesis produces oxygen that lacks mass-independent fractionation, so higher rates of O$_2$ production (e.g. smaller $\tau$) will tend to wash out the stratospheric fractionation signal in the troposphere. That is why $\rho$, $\gamma$, $\Theta$, and $\tau$ all appear in the numerator. $\rho$ appears in the denominator because very large $\rho$ (e.g. pO$_2 \gg$ pCO$_2$) limits the fraction of heavy oxygens from O$_2$ that can be sequestered in CO$_2$ molecules. $\gamma$, $\Theta$, and $\tau$ appear in the denominator because the magnitude of the tropospheric $\Delta^{17}$O anomaly cannot exceed the value reached in the upper stratosphere, so there must be a limit past which more efficient transport of altered oxygen from the upper stratosphere to the troposphere (e.g. increases in $\gamma$ or $\Theta$) or reduced photosynthetic flux of unaltered oxygen (e.g. increases in $\tau$) stop affecting the troposphere's value. We focus on estimating $\gamma$ and $\Theta$ in this study.}

\subsubsection{Stratosphere-troposphere exchange}\label{sec:ste}

The cross-tropopause mass exchange rate $\gamma$ can be represented as the ratio between the flux of mass across the tropopause and the total mass of the atmosphere, giving a quantity with units of yr$^{-1}$. The $\gamma$ used in the snowball calculations in \citet{Bao09,Cao2013} to infer high CO$_2$ during the snowball is derived from data in \citet{appenzeller1996seasonal}, which describes the modern atmosphere. {We directly calculate the upward mass flux across the tropopause in our simulations by taking the difference between the sum of all local maxima and the sum of all local minima of the STEM circulation along an isentrope that roughly coincides with the highest portion of the tropopause in potential temperature coordinates (380 K in the modern simulation, and 285 K in the snowballs)}:
\begin{linenomath}
\begin{align}
\gamma &= \frac{{\sum}\Psi_{\text{STEM}}(\theta_{\text{tropo}})_{\text{max}}-{\sum}\Psi_{\text{STEM}}(\theta_{\text{tropo}})_{\text{min}}}{M_{\text{atm}}}\\
&=\frac{F_{\text{STE}}}{M_{\text{atm}}}
\end{align}
\end{linenomath}
where $F_{\text{STE}}$ is the mass flux across the tropopause, $\theta_{\text{tropo}}$ is potential temperature chosen to represent the tropopause, and $M_{\text{atm}}$ is the mass of the atmosphere as determined by the pressure it exerts at Earth's surface. The globally averaged surface pressure in each of the simulations is $9.855\times 10^{4}$ Pa. {Here, the surface pressure refers to the actual pressure on the ground, including the effects of topography, and not the mean sea level pressure.} $M_{\text{atm}}$ is calculated by dividing the surface pressure by the gravity $g$. {We calculate the mass flux across the isentrope that coincides with the highest portion of the tropopause in our simulations (instead of directly across the tropopause) because those potential temperatures approximately coincide with the bottom of the atmospheric region in which isentropes fall entirely within the stratosphere, e.g. the ``stratospheric overworld'' \citep{holton1995stratosphere}. There is a large amount of adiabatic mass exchange with the troposphere along the isentropes in the ``lowermost stratosphere'' that cross the tropopause. Since that portion of the stratosphere is more tightly coupled to the troposphere by mass exchange than it is to the ``overworld'' \citep{holton1995stratosphere}, it is better considered a part of the troposphere for the purpose of calculating $\Delta^{17}$O.}

\subsubsection{Stratospheric mixing efficiency}\label{Theta}
The stratospheric mixing efficiency, denoted $\Theta$ in this study, is defined in \citet{Cao2013} as the ``fraction of the steady-state O$_2$ in the O$_2$\textendash{}$\text{CO}_2${\textendash{}} O$_3$ reaction network within the total O$_2$ in the stratosphere.'' In \citet{Cao2013}, $\Theta$ is inferred from equation (\ref{eq_delta_17}) using modern values of $\Delta^{17}$O (-0.34$\permil$), pCO$_2$ (375 ppmv), pO$_2$ (209,500 ppmv), and $\tau$ (1244 years). To examine whether this parameter may have been different during a snowball, changing estimates of pCO$_2$ made using the $\Delta^{17}$O values presented in \citet{Bao09}, we estimate $\Theta$ using data from our simulations. 

ECHAM6 does not contain interactive chemistry, and there is no standard formula to directly calculate $\Theta$. To estimate $\Theta$, we calculate the fraction of mass at a given altitude that has passed through the region of the stratosphere where rates of photochemistry would peak, but has not yet passed through the troposphere, and we take $\Theta$ to be directly proportional to this quantity at the tropopause. We call this quantity $\xi$ {and treat it as the concentration of a passive tracer}, and we use ``A'' to denote the scaling factor that relates $\xi(\text{tropopause})$ to $\Theta$, such that $\Theta = \text{A}\times\xi(\text{tropopause})$. Assuming $\gamma\times\Theta$ from our modern simulation to be equal to the $\gamma\times\Theta$ given in \citet{Cao2013} allows us to calculate the proportionality constant A, since in that case $\text{A} = \frac{\gamma_{\text{Cao}}\times\Theta_{\text{Cao}}}{\gamma\times\xi}$. We can then use this value of A and the $\xi$ values calculated from our snowball simulations to estimate how $\Theta$ changes under snowball conditions.

{We assume that the altitude where the prescribed O$_3$ peaks in concentration is the region of the stratosphere where isotopically altered O$_2$ is produced, since this process is ultimately due to UV photolysis of ozone (see discussion above in section \ref{sec:delta_17}).} The potential temperature of this region is 1000 K in each of the simulations with ozone, so we use this value as the ``source'' altitude. {We do not account for the fact that the polar night region of the stratosphere would not be photochemically active, since less than 10$\%$ of the stratosphere is in polar night at a given time, and this is a rough estimate of $\Theta$.} The same value is used to calculate $\Theta$ in the simulation without ozone for ease of comparison, despite the fact that a stratosphere lacking ozone would not produce this fractionation signal. We use the no ozone simulation as a limiting case to study the properties of stratospheres with reduced ozone abundances. We determine the ``sink'' altitude's potential temperature in each case by calculating the spatial and temporal average of the potential temperature at the tropopause. The rate of change of $\xi$ in a region with a given potential temperature $\theta$ can be calculated by multiplying each mass flux into and out of that region by the $\xi$ of the region the flux comes from, giving the fluxes of ``photochemically altered'' mass into and out of the region with that potential temperature from the regions directly above and below it. This allows us to write down a conservation equation at each potential temperature:
\begin{linenomath}
\begin{align}
    \frac{d\xi (\theta)}{dt} &= F_{\text{above}}\times\xi(\theta + d\theta)-F_{\text{above}}\times\xi(\theta)+ F_{\text{below}}\times\xi(\theta-d\theta) - F_{\text{below}}\times\xi(\theta)
\end{align}
\end{linenomath}
where $F_{\text{above}}$ is the flux of mass between stratospheric regions with potential temperatures $\theta$  and $\theta + d\theta$ and $F_{\text{below}}$ is the flux of mass between regions with potential temperatures $\theta$ and $\theta-d\theta$. Each $F$ is calculated as the difference between the maximum and minimum of the STEM circulation at the boundary between two regions. {Figure \ref{fig:xi_schematic} shows a schematic of a stratospheric layer with potential temperature $\theta$ exchanging mass with layers with potential temperatures $\theta+d\theta$ and $\theta-d\theta$ via $F_\text{above}$ and $F_\text{below}$.} Assuming steady-state, we can then derive an equation for $\xi$:
\begin{linenomath}
\begin{align}
    \xi(\theta) &= \frac{F_{\text{above}}\times\xi(\theta + d\theta) + F_{\text{below}}\times\xi(\theta-d\theta)}{F_{\text{above}}+F_{\text{below}}} \label{eq_xi}
\end{align}
\end{linenomath}
 We solve this system of equations (equation (\ref{eq_xi}) for each zonal- and time-averaged potential temperature) using matrix inversion, with the boundary conditions that $\xi$ is equal to zero immediately below the tropopause (because $\xi$ is a measure of the fraction of mass that has reached a given point directly from the upper stratosphere without passing through the troposphere) and $\xi$(1000 K) -- where the prescribed ozone peaks -- is equal to one. Multiplying the value of $\xi$ at the tropopause by the scaling factor A gives $\Theta$.

\section{Results}
\label{results}

We begin by applying the STEM circulation, including its extension with advection, eddy momentum flux convergence, and diffusion, to reanalysis data and the modern simulation. Next, we quantify changes in the stratospheric circulation between the modern and snowball simulations. We then connect the circulation changes to changes in wave driving, the stratospheric polar vortex, sudden stratospheric warmings, and stratospheric temperatures. Finally, we compare the values of $\gamma$ and $\Theta$ under modern and snowball conditions.

\subsection{Stratospheric circulation}
\label{strat_circ}

\begin{figure}
\vspace*{2mm}
\makebox[\textwidth][c]{\includegraphics[width=50pc]{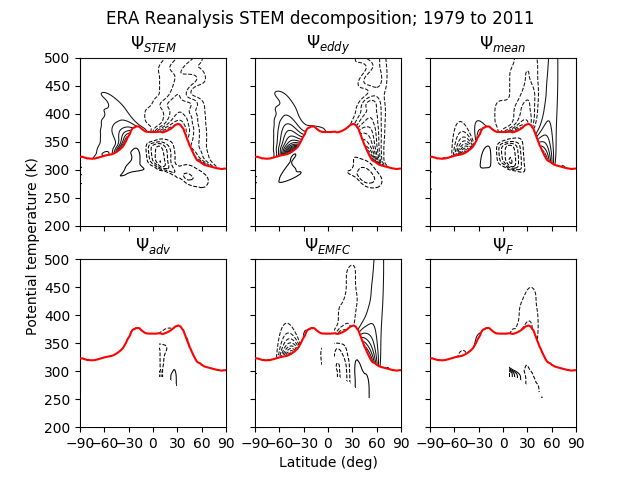}}
\caption{Decomposition of the statistical transformed Eulerian-mean stream function of daily ERA-interim reanalysis data from 1979 to 2011. The stream functions shown are $\Psi_{\text{STEM}}$, $\Psi_\text{{eddy}}$, $\Psi_\text{{mean}}$, $\Psi_\text{adv}$, $\Psi_\text{EMFC}$, and $\Psi_{\text{{F}}}$ (see text for details). $\Psi_\text{adv}$, $\Psi_\text{EMFC}$, and $\Psi_\text{{F}}$ are whited out between -{8.4} and {8.4} degrees because they each have a Coriolis term in the denominator, which leads to inaccurate behavior near the equator. The vertical axis is potential temperature averaged over December, January, and February. The contour interval is $4.0\times10^{10} \frac{kg}{s}$ below the tropopause averaged in the zonal mean and over December, January, and February (represented by the red curve) and $2.0\times10^9 \frac{kg}{s}$ above it. The zero contour is not shown. Solid contours indicate counter-clockwise flow, and dashed contours indicate clockwise flow.}
\label{fig:era}
\end{figure}

\begin{figure}
\vspace*{2mm}
\makebox[\textwidth][c]{\includegraphics[width=50pc]{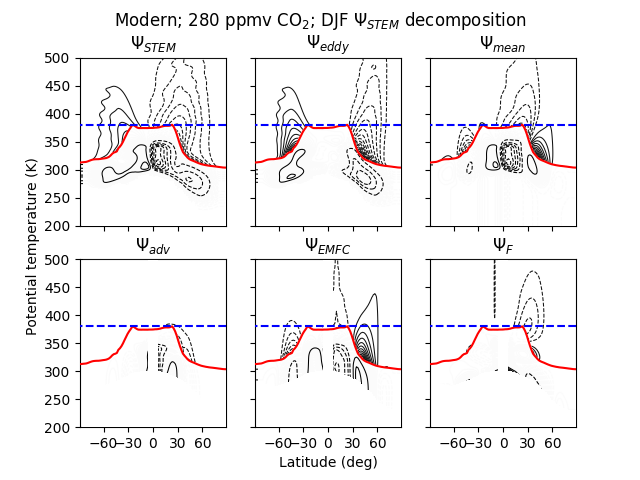}}
\caption{Same as Fig.\ref{fig:era}, but for our modern simulation. The dotted blue line is at 380 K, the isentrope we used to mark the stratospheric ``overworld'' in the modern simulation (see text).}
\label{fig:modern}
\end{figure}

\begin{figure}
\vspace*{2mm}
\makebox[\textwidth][c]{\includegraphics[width=35pc]{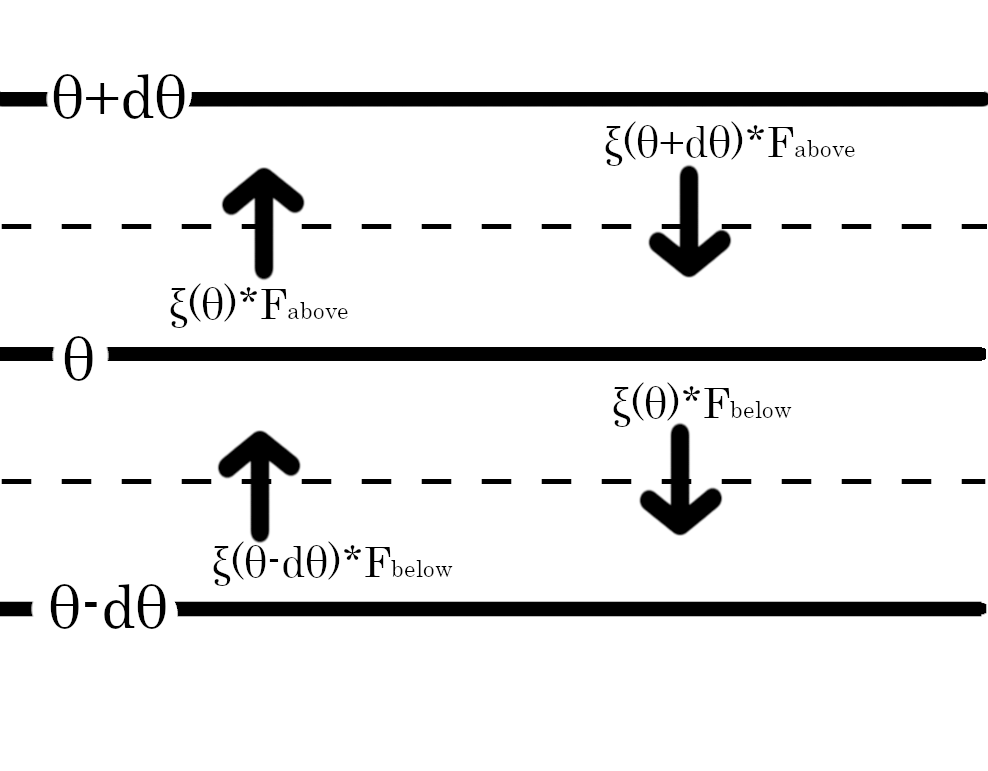}}
\caption{{A schematic of the one-dimensional discretization used to calculate $\xi$ at each potential temperature in the stratosphere. The solid black lines mark the center points of the regions in the stratosphere with spatially- and temporally-averaged potential temperatures of (from top) $\theta+d\theta$, $\theta$, and $\theta-d\theta$. The dashed lines represent the boundaries between those regions. $F_\text{above}$ is the mass exchange between the region with potential temperature $\theta+d\theta$ and the region with potential temperature $\theta$. $F_\text{below}$ is the mass exchange between region with potential temperature $\theta$ and the region with potential temperature $\theta-d\theta$. The mass fluxes are each multiplied by the $\xi$ of the region they are leaving to give the fluxes of tracer between regions.}}
\label{fig:xi_schematic}
\end{figure}

Since this is the first time the STEM framework and its extension have been used to analyze the stratospheric circulation, we apply it to the ERA-Interim reanalysis dataset (see e.g. \citet{dee2011era} for discussion of the ERA-Interim data; see Fig. \ref{fig:era} for our decomposition of the data). The full STEM circulation, $\Psi_{\text{STEM}}$, successfully reproduces the dry isentropic circulation in the stratosphere and all the features of the TEM stratospheric circulation (see Figure 5 in \citet{butchart2014brewer}). Further, through the decomposition into different dynamical components (see equations {\ref{eqn:psi_adv} to \ref{eqn:psi_diff}}), we see that the stratospheric circulation is largely driven by eddies (see Fig. \ref{fig:era}, $\Psi_\text{eddy}$ {$\&$} {$\Psi_\text{EMFC}$}), as expected \citep{butchart2014brewer}. {Unresolved processes, possibly gravity wave drag or analysis increments associated with the data assimilation scheme, are responsible for a non-negligible portion of the stratospheric circulation between 0$^{\circ}$ and 30$^\circ$} (see Fig. \ref{fig:era}, $\Psi_\text{{F}}$). We see similar results for the modern simulation (Fig. \ref{fig:modern}). The simulation displays a slightly stronger atmospheric circulation than the ERA-Interim reanalysis data because it lacks ocean heat transport (see e.g. \citet{czaja2006partitioning}).

\begin{figure}
\vspace*{2mm}
\makebox[\textwidth][c]{\includegraphics[width=50pc]{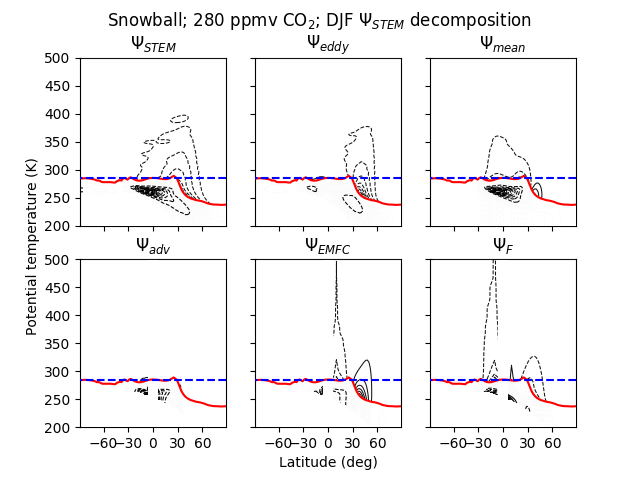}}
\caption{Same as Fig. \ref{fig:modern}, but for the 280 ppmv CO$_2$ snowball simulation with modern ozone and topography. The dotted blue line is at 285 K, the isentrope we used as the bottom boundary to the stratospheric ``overworld'' in the snowball simulations (see text).}

\label{fig:280ppmv}
\end{figure}

\begin{figure}
\vspace*{2mm}
\makebox[\textwidth][c]{\includegraphics[width=50pc]{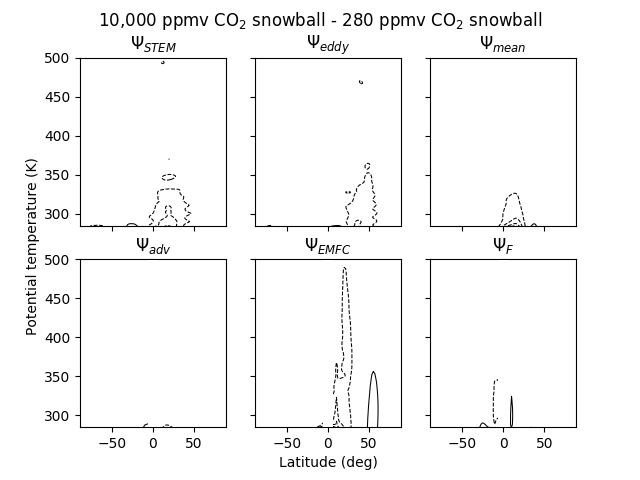}}
\caption{{Statistically significant differences between the stratospheric circulations of the 10,000 ppmv CO$_2$ snowball simulation and the 280 ppmv CO$_2$ snowball simulation with modern ozone and topography. The contour interval is $8\times10^8\frac{kg}{s}$. The bottom boundary in these plots is the 285 K isentrope that approximately separates the ``stratospheric overworld'' from the ``lowermost stratosphere'' in our snowball simulations (see Section \ref{sec:ste}).}}
\label{fig:10000ppmv_difference}
\end{figure}

\begin{figure}
\vspace*{2mm}
\makebox[\textwidth][c]{\includegraphics[width=50pc]{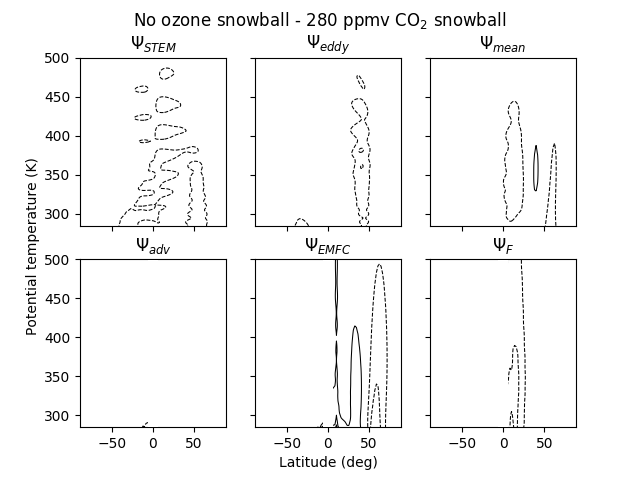}}
\caption{{Statistically significant differences between the stratospheric circulations of the snowball simulation without ozone and the 280 ppmv CO$_2$ snowball simulation with modern ozone and topography. The contour interval is $8\times10^8 \frac{kg}{s}$. The bottom boundary in these plots is the 285 K isentrope that approximately separates the ``stratospheric overworld'' from the ``lowermost stratosphere'' in our snowball simulations (see Section \ref{sec:ste}).}}
\label{fig:nozone_difference}
\end{figure}

\begin{figure}
\vspace*{2mm}
\makebox[\textwidth][c]{\includegraphics[width=50pc]{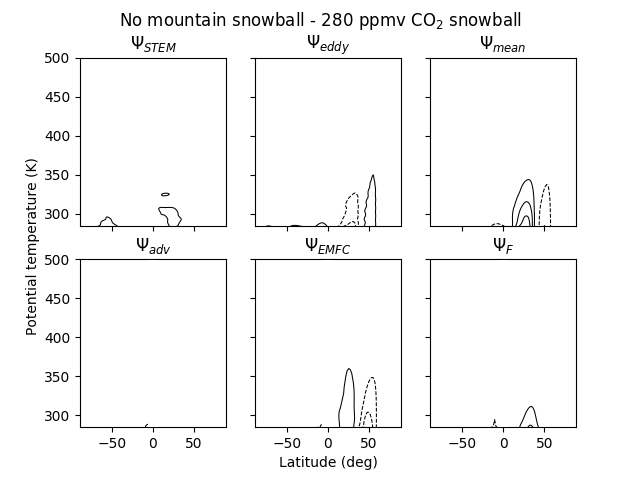}}
\caption{{Statistically significant differences between the stratospheric circulations of the snowball simulation with flattened topography and the 280 ppmv CO$_2$ snowball simulation with modern ozone and topography. The contour interval is $8\times10^8\frac{kg}{s}$. The bottom boundary in these plots is the 285 K isentrope that approximately separates the ``stratospheric overworld'' from the ``lowermost stratosphere'' in our snowball simulations (see Section \ref{sec:ste}).}}
\label{fig:nomount_difference}
\end{figure}

Consistent with previous work by \citet{Pierrehumbert05,Voigt2012,voigt2013dynamics,abbot13-snowball-circulation}, the {winter-hemisphere} Hadley cell is much stronger in the snowball simulations than in the modern case (Tables \ref{tab:sims}, \ref{tab:psi_comps_winter}; Figs. \ref{fig:modern} {$\&$ \ref{fig:280ppmv}}). In contrast, the stratospheric circulation is notably weaker in both summer and winter (Tables \ref{tab:sims}, \ref{tab:psi_comps_winter}, \ref{tab:psi_comps_summer}; Figs. \ref{fig:modern} {$\&$ \ref{fig:280ppmv}}). Increasing CO$_2$ from 280 ppmv to 10,000 ppmv in the snowball simulation strengthens the stratospheric flow (Fig. \ref{fig:10000ppmv_difference}). {Removing ozone in the snowball simulation increases the depth of the stratospheric circulation (Fig. \ref{fig:nozone_difference}) and raises the height of the tropopause, calculated according to the standard WMO definition}. Flattening the topography further weakens the snowball stratosphere's circulation and shifts {$\Psi_{\text{EMFC}}$} to lower altitudes compared to the mountainous simulations (Fig. \ref{fig:nomount_difference}).

Largely consistent with one of the main results of \citet{Voigt2012,voigt2013dynamics}, $\Psi_\text{{F}}$ is a major driver of $\Psi_\text{mean}$ in the troposphere of our snowball simulations, accounting for a substantial portion in the winter cell, and nearly the entire mean circulation in the summer cell (see Tables \ref{tab:psi_comps_winter} \& \ref{tab:psi_comps_summer}; the summer cell is too weak to be visible in the snowball circulation figures). The situation is similar in the winter stratosphere, with $\Psi_\text{{F}}$ accounting for between $22\%$ and $80\%$ of the strength of $\Psi_\text{STEM}$ at its extremum (Table \ref{tab:psi_comps_winter}). $\Psi_\text{eddy}$ is also a major driver of the winter snowball stratosphere's circulation, especially poleward of 30$^{\circ}$ (see Fig. \ref{fig:280ppmv}), but specifically $\textit{at the point where}$ $\Psi_\text{STEM}$ $\textit{reaches}$ $\textit{its}$  $\textit{maximum}$ $\textit{in}$ $\textit{the}$  $\textit{stratospheric}$ $\textit{overworld}$, $\Psi_\text{{F}}$ plays a large role. In the summer hemisphere snowball stratosphere, $\Psi_\text{STEM}$ is dominated by $\Psi_\text{eddy}$, but $\Psi_\text{{F}}$ is the main component of $\Psi_\text{mean}$ (the summer circulation is too weak to be seen in the figures of snowball stratospheric circulation at the contour interval we used; see Table \ref{tab:psi_comps_summer}). $\Psi_\text{{F}}$ seems to be responsible for some of the highest-altitude transport that occurs in the snowball stratosphere (Fig. \ref{fig:280ppmv}).

\begin{figure}
\vspace*{2mm}
\makebox[\textwidth][c]{\includegraphics[width=38pc]{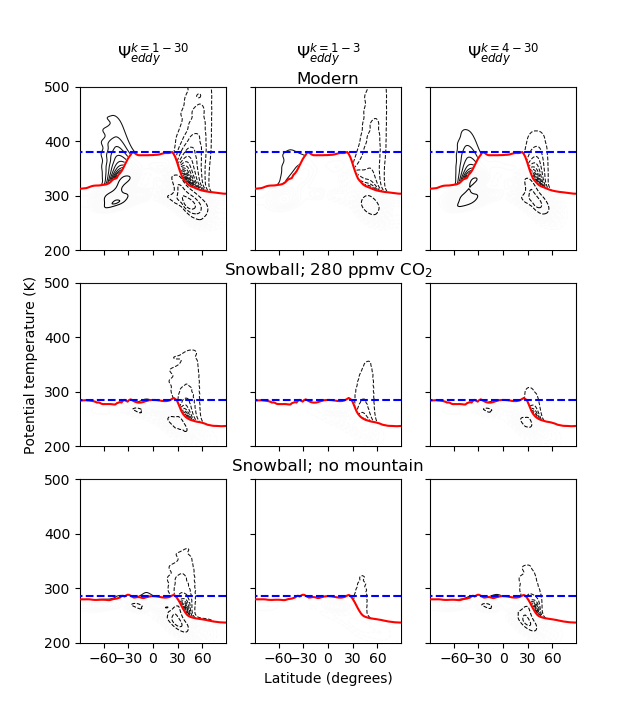}}
\caption{A wavenumber decomposition of $\Psi_\text{eddy}$. The top row of plots shows the decomposition for the modern simulation, the middle shows it for the 280 ppmv CO$_2$ snowball simulation with modern ozone and topography, and the bottom shows the results from the mountainless snowball simulation. The {middle} column of plots shows the circulation driven by planetary-scale (wavenumber k = 1 -3) waves, the {right} column shows the circulation driven by synoptic-scale (wavenumber k=4-30) waves, and the {left} column shows their sum. The contour interval is $2\times10^9\frac{\text{kg}}{\text{s}}$ above the tropopause and $4\times10^{10}\frac{\text{kg}}{\text{s}}$ {below the tropopause.}}
\label{fig:wvn_decomp}
\end{figure}

Most of the reduction in the strength of the snowball stratosphere's circulation can be attributed to changes in the wave-driven components. Planetary-scale eddy-driven circulation is strongly reduced in the 280 ppmv CO$_2$ snowball with modern ozone and topography compared to the modern simulation (see Fig. \ref{fig:wvn_decomp}). This is likely at least partly due to a reduction in planetary-scale Rossby waves because of the lack of land-sea contrast on a fully ice-covered snowball. The flattened snowball simulation displays even weaker planetary-scale wave-driven circulation because of a lack of Rossby waves that are generated as wind blows over topography (Fig. \ref{fig:wvn_decomp}). It is also clear from Fig. \ref{fig:wvn_decomp} that planetary-scale eddies contribute to the ``deep branch'' of the stratospheric circulation and synoptic-scale eddies to the ``shallow branch'' in both modern and snowball simulations. The synoptic-scale component of $\Psi_\text{eddy}$ also weakens in the snowball simulations, particularly in the summer hemisphere, although not by as much as the planetary component. The ``shallow'' synoptic-scale component of the flattened snowball's circulation is more vigorous than that of the snowball with modern topography (Fig. \ref{fig:wvn_decomp}).


\subsection{Wave activity}
\label{strat_wave_act}

When considering the polar {winter} stratosphere as a box, the reduction in upward wave propagation in the snowball relative to modern leads to decreased EP flux convergence in the box because the flux out of the subtropical boundary is mostly unchanged in the simulations with ozone (Fig. \ref{fig:ep_conv}). The simulation without ozone displays a similar convergence to the other snowball simulations despite considerably weaker influx across the bottom boundary because its subtropical boundary outflux is also strongly reduced. Since the stratospheric circulation is driven by the EP flux convergence, a reduction of vertical EP flux with small changes in meridional EP flux weakens the circulation as shown in Fig. \ref{fig:wvn_decomp}.

\subsection{Stratospheric jets}
\label{strat_jets}
{The stratospheric polar vortex occurs due to the potential temperature gradient generated by solar heating of stratospheric ozone in the tropics and strong radiative cooling by CO$_2$ in the polar night region, and the strength of the vortex is modulated by the convergence of EP flux \citep{waugh2010stratospheric}. Because EP flux convergence acts as a torque on the zonal-mean flow and decelerates it, the weakened convergence over the polar cap in the snowballs with modern ozone is consistent with a strengthened polar vortex compared to the modern simulation (right column of Fig. \ref{fig:jets}). In contrast, the snowball simulation without ozone does not display a vortex, since it lacks the temperature gradient generated by ozone heating.}

\begin{figure}
\vspace*{2mm}
\begin{center}
  \includegraphics[width=35pc]{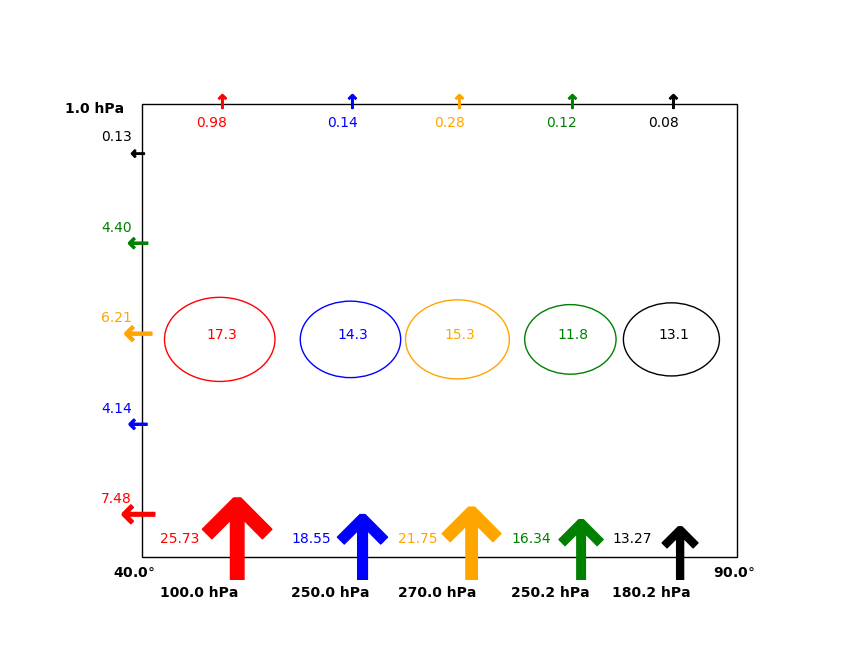}
\end{center}
\caption{This plot shows the convergence of the Eliassen-Palm (EP) flux vector integrated over the boundaries of a box with sides enclosed by latitudes of $40.0 \degree \leq \phi \leq 90.0 \degree$ and pressures of p$_1$ $\geq$ p $\geq$ $1.0$ hPa, where p$_1$ is the area-weighted average pressure of the isentrope used as the tropopause for each simulation. The total convergences are represented by the circles in the center for the modern simulation (red), the 280 ppmv CO$_2$ snowball with modern ozone and topography (blue), the 10,000 ppmv CO$_2$ snowball (orange), the 280 ppmv CO$_2$ snowball with flattened topograph (green), and the 280 ppmv CO$_2$ snowball with no ozone (black). The arrows indicate the direction and magnitude of the integrated wave activity flux across each boundary. No wave flux enters or leaves the $90.0 \degree$ boundary.}
\label{fig:ep_conv}
\end{figure}

\begin{figure}
\vspace*{2mm}
\begin{center}
  \includegraphics[width=35pc]{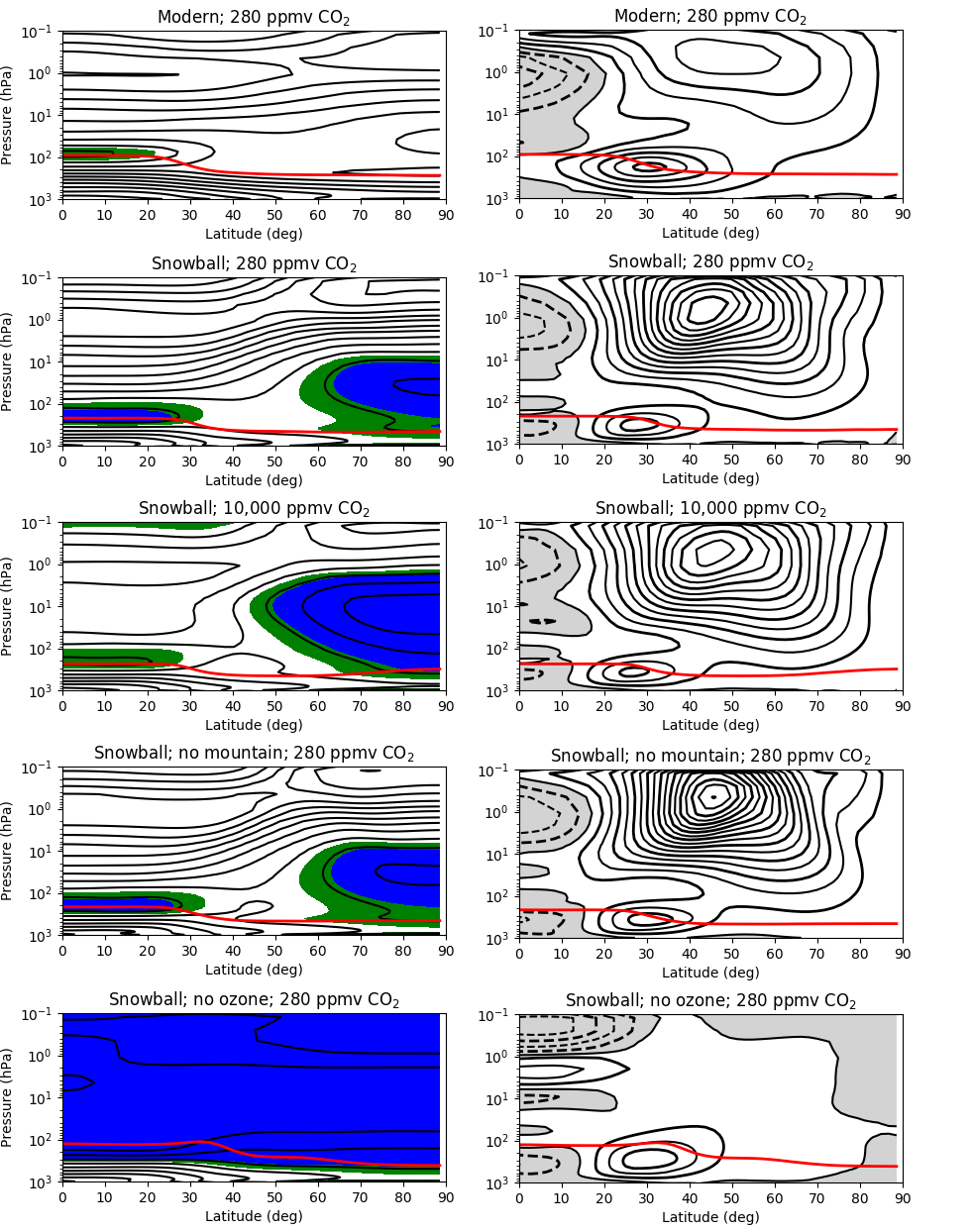}
\end{center}
\caption{Left: The zonal-mean atmospheric temperatures averaged over December, January, and February (DJF), for, from top to bottom, the modern simulation, the 280 ppmv CO$_2$ snowball simulation with modern ozone and topography, the 10,000 ppmv CO$_2$ snowball simulation, the 280 ppmv CO$_2$ snowball simulation with flattened topography, and the 280 ppmv CO$_2$ snowball simulation with no ozone. The contour interval is 10 K. Green regions mark where the temperature falls below 195 K, roughly that required for Type I polar stratospheric cloud (PSC) formation. Blue regions mark temperatures below 188 K, at which Type II PSCs are observed to form. The {red} lines mark the tropopause averaged over DJF. Right: The zonal-mean jet speeds averaged over DJF for the same simulations. The contour interval is 10 m s$^{-1}$. The grey zones represent easterly winds, while the unshaded zones represent westerlies.}
\label{fig:jets}
\end{figure}

The differences in jet speeds among the three snowball simulations with modern ozone are also consistent with the modulation of vortex strength by EP flux convergence. The small, 1 kg m s$^{-4}$ increase in EP flux convergence between the 280 ppmv and 10,000 ppmv CO$_2$ snowball simulations (compare the blue and orange circles in  Fig. \ref{fig:ep_conv}) mirrors a small, $\sim$10 m s$^{-1}$ reduction in jet speed (Fig. \ref{fig:jets}). The larger 2.5 kg m s$^{-4}$  decrease in EP flux convergence between the 280 ppmv CO$_2$ snowball simulation and the no mountain snowball simulation (compare the blue and green circles in Fig. \ref{fig:ep_conv}) accompanies a 30 m s$^{-1}$ increase in jet speed (Fig. \ref{fig:jets}). 

EP flux convergence over the polar cap is also responsible for creating sudden stratospheric warming (SSW) events where the zonal-mean zonal wind reverses sign through deceleration of the polar vortex \citep{butler2015defining}. In \citet{charlton2007new}, an event is defined to occur when the zonal-mean zonal winds at a latitude of $60\degree$ and a pressure of 10 hPa become easterly during the period of November to March. At least twenty days must pass after an event before another can be identified, and winds must return to westerly for at least ten days prior to April 30th to avoid classification as a ``final warming.'' {\citet{charlton2007new} find that the modern Earth experiences approximately 0.62 SSWs per year, while our modern simulation produces about one SSW per year, suggesting that the simulation has a bias toward weak zonal winds. The frequency of SSWs goes from about one per year for the modern simulation to none at all for the snowball simulations, which is striking, given the bias in the modern simulation.} However, with only 5 years of simulation data, we cannot rule out the possibility that SSWs might occur at low frequencies in the snowball simulations. {This is similar to the situation in the modern Earth's Southern Hemisphere stratosphere, where weak wave driving leads to a polar vortex that is stronger than in the Northern Hemisphere, with very infrequent SSWs \citep{waugh2010stratospheric}.}
\subsection{Stratospheric Temperature}

The weakening of the stratospheric circulation, vertical EP flux, and  EP flux convergence as well as the strengthening of the stratospheric polar vortex all imply colder stratospheric temperatures. In particular, in the absence of strong adiabatic warming induced by a vigorous stratospheric circulation, we would expect the temperatures to be closer to radiative equilibrium. This is indeed the case in the snowball simulations with modern ozone (left column of Fig. \ref{fig:jets}). At any given time more than 65$\%$ of the Earth's surface in the snowball simulations with modern ozone is beneath stratospheric air that has a temperature less than 195K, the threshold for type 1 polar stratospheric cloud formation \citep{kinne1990radiative}. In contrast, only around $30\%$ of the modern simulation's surface lies beneath stratospheric air colder than 195K, and none falls below 188 K, the threshold for type 2 polar stratospheric cloud formation \citep{kinne1990radiative}, except occasionally in the {Antarctic} during polar winter. In the snowball simulation without ozone, the entire stratosphere falls below 188 K because there is no solar heating of ozone to warm it.

\subsection{Cross-tropopause mass exchange rate, stratospheric mixing efficiency, and $\Delta^{17}$O}

The cross-tropopause mass exchange rate $\gamma$ of 0.12 yr$^{-1}$ calculated from our modern simulation (Table \ref{tab:sims}) is close to the $\gamma$ of 0.13 yr$^{-1}$ calculated from data in \citet{appenzeller1996seasonal} and used in \citet{Bao08,Bao09,Cao2013}. In turn, the modern $\Theta$ of 0.018 is similar to the value of 0.017 used by Cao and Bao (by construction, see section \ref{Theta})). Each snowball simulation has a $\gamma$ that is smaller than that of the modern simulation, while each snowball simulation has a $\Theta$ value that is larger than that of the modern simulation (Table \ref{tab:sims}). 

\begin{figure}
\vspace*{2mm}
\makebox[\textwidth][c]{\includegraphics[width=50pc]{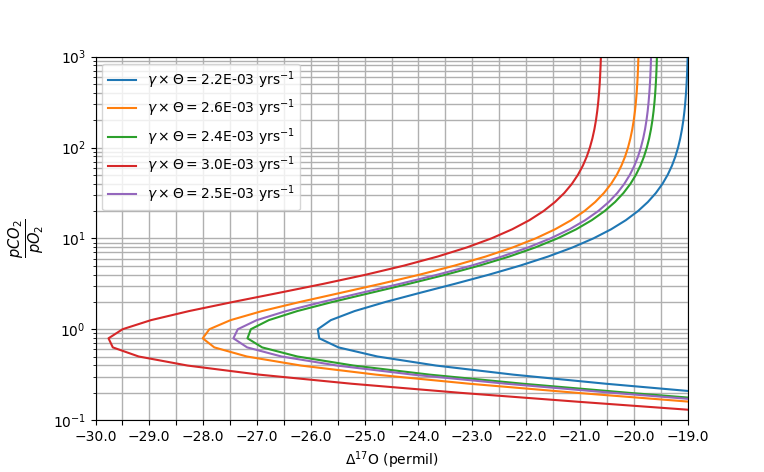}}
\caption{Tropospheric $\Delta^{17}$O in $\permil$ vs $\frac{\text{pCO}_2}{\text{pO}_2}$, varying $\gamma$ $\times$ $\Theta$ over the values from Table \ref{tab:sims} to see how incorporating results from snowball simulations impacts CO$_2$ inferences from $\Delta^{17}$O measurements. Values in Equation (\ref{eq_delta_17}) other than $\gamma$ and $\Theta$ are taken from \citet{Cao2013}. The blue line uses a $\gamma \times \Theta$ equal to that of the modern simulation, orange corresponds to the value from the 280 ppmv CO$_2$ snowball with modern ozone and topography, green corresponds to the value from the 10,000 ppmv CO$_2$ snowball, red corresponds to the value from the snowball without ozone, and purple corresponds to the value from the snowball without mountains.}
\label{fig:CO2_and_gammatheta}
\end{figure}

\citet{Bao08,Bao09,Cao2013} estimated the CO$_2$ concentration during and after the snowball from Neoproterozoic $\Delta^{17}$O by assuming modern $\gamma$ and $\Theta$. {Since $\Delta^{17}$O is a function of the product of $\gamma$ and $\Theta$ (see equation (\ref{eq_delta_17})) the decreases in $\gamma$ and increases in $\Theta$ between the modern simulation and snowball simulations end up partially canceling. The largest deviation from modern in $\gamma\times\Theta$ is displayed by the snowball simulation without ozone, with a value $27\%$ larger than modern.} To see how the relationship between $\frac{\text{pCO}_2}{\text{pO}_2}$ and $\Delta^{17}$O depends on $\gamma$ and $\Theta$ calculated from our simulations, we vary $\gamma\times\Theta$ across the values in Table \ref{tab:sims} while holding the other variables in equation (\ref{eq_delta_17}) constant at the values used in \citet{Bao08,Bao09} and listed in Table 1 of \citet{Cao2013} (see our Fig. \ref{fig:CO2_and_gammatheta}). 

{The variations in $\gamma\times\Theta$ can, depending on the tropospheric $\Delta^{17}$O value to be explained, produce a maximum increase in estimated CO$_2$ of up to $91\times$ or a maximum decrease by a factor of 3.6, measured relative to estimates produced with the modern $\gamma\times\Theta$.} However, the region in Fig. \ref{fig:CO2_and_gammatheta} that displays the largest variation in $\frac{\text{pCO}_2}{\text{pO}_2}$ as a function of $\gamma\times\Theta$ is also a region where pCO$_2$ is orders of magnitude greater than pO$_2$, which is arguably not likely to have been the case during the Neoproterozoic. \citet{canfield1996late,holland2006oxygenation,kump2014hypothesized,laakso2017theory} all estimate that Neoproterozoic oxygen levels may have been equal to or greater than 10$\%$ of modern pO$_2$, implying an atmospheric concentration $\geq$20,000 ppmv, and \citet{kasemann2005boron} use boron and calcium isotopes to estimate a maximum late Neoproterozoic CO$_2$ concentration of 90,000 ppmv. In the region of the parameter space where pCO$_2 \leq$ 10$\times$pO$_2$, the values of $\frac{\text{pCO}_2}{\text{pO}_2}$ remain close together over the range of $\gamma\times\Theta$ produced by our simulations, which helps validate the use of modern $\gamma$ and $\Theta$ to estimate snowball CO$_2$ in \citet{Bao08,Bao09,Cao2013}. 


\section{Conclusion $\&$ Discussion}
\label{discussion}
\subsection{Conclusions}
In this study we analyzed GCM simulations of the stratosphere of the snowball Earth. We focused on changes in the stratospheric circulation, EP flux convergence, polar vortex, and temperatures relative to modern conditions. The stratospheric circulation was examined using the Statistical Transformed Eulerian Mean framework of \citet{pauluis2011statistical}. We extended the STEM framework to include components accounting for advection, eddy momentum flux convergence, and diffusion, since \citet{Voigt2012,voigt2013dynamics} demonstrated the importance of diffusion for the snowball Hadley cell.


We found that the stratospheric circulation during a snowball would have been weakened compared to the modern. The reduced circulation strength in the winter stratosphere of the Northern Hemisphere was linked to a weakening of the torque induced by eddy convergence in the stratosphere. The decreased torque was likely caused primarily by a reduction in Rossby waves due to the lack of land-sea contrast in the snowball simulations. In the snowball simulations, wave activity at both planetary and synoptic scales was generally weakened relative to modern. The weakening of eddy convergence led to a much stronger polar vortex in snowball simulations with modern ozone, while the lack of differential stratospheric solar heating in the case without ozone eliminated the vortex entirely. Reduced eddy convergence also eliminated sudden stratospheric warming events in all snowball simulations. In the simulations with ozone, the weak eddy heat flux led to extremely low stratospheric temperatures, particularly over the poles where the powerful vortex isolated the stratospheric air. Since the simulation without ozone had much less UV absorption, the temperatures in the stratosphere were even lower.
 
The anomalous $\Delta^{17}$O signal discovered in Neoproterozoic sulfates and interpreted as evidence for elevated CO$_2$ during a snowball in \citet{Bao09} based on the model presented in \citet{Cao2013} is dependent on the product of the stratosphere-troposphere exchange ($\gamma$) and the stratospheric mixing efficiency ($\Theta$). In this study, we quantified $\gamma$ and $\Theta$ under snowball conditions from GCM output for the first time. {We showed that reductions in $\gamma$ and increases in $\Theta$ under snowball conditions partially cancel one another in our simulations. This makes intuitive sense, as a reduction in the flux of material from the troposphere into the stratosphere (a smaller $\gamma$) should make it easier to homogenize the isotopic composition of the stratosphere (an increase in $\Theta$) and vice versa. This is most clearly illustrated by considering the limit where $\gamma$ is zero--the composition of the stratosphere would homogenize without a flux of material from the troposphere, which is equivalent to having a $\Theta$ of one (perfect stratospheric mixing efficiency). The use of modern values for those variables in \citet{Bao08,Bao09,Cao2013} should not introduce large errors in estimates of snowball CO$_2$ from $\Delta^{17}$O, unless there was orders of magnitude more CO$_2$ than O$_2$ in the snowball atmosphere.} There are larger uncertainties in the estimate of pCO$_2$ from $\Delta^{17}$O due to the lack of information about what fraction of oxygen in the mineral samples came from the atmosphere (see \citet{Bao08,Bao09}). 
\subsection{Discussion}
Because our models did not have interactive chemistry, the implications of our findings for $\Delta^{17}$O are tentative. The photochemistry that generates the fractionation signal is determined by the temperatures, bulk velocities, and spatial distributions of participating O$_3$, O$_2$, and CO$_2$ molecules. Given the extreme cold, powerful jet (in the modern ozone cases) or weakened jet (in the no ozone case, which can be considered an end-member of reduced ozone scenarios), and weakened Brewer Dobson circulation, a detailed quantification of the effects of changes to these properties on estimates of the magnitude of $\Delta^{17}$O is warranted. This will better constrain how CO$_2$ concentration influenced oxygen fractionation under the exotic conditions of the snowball. Chemical transport models would be useful in exploring the evolution of the coupled O$_3$-O$_2$-CO$_2$ photochemical network during the Neoproterozoic glaciations.

The extremely cold temperatures in the stratospheres of our snowball simulations might allow for the formation of ``polar stratospheric clouds'' (PSCs) far from the poles (see e.g. \citet{steele1983formation, kinne1990radiative} for discussion of PSCs). If present, optically thick PSCs would provide a strong greenhouse effect \citep{Sloan-Pollard-1998:polar,Sloan-Huber-Ewing-1999:polar, Kirk-Davidoff-Schrag-Anderson-2002:feedback}. If conditions allowed for the formation and persistence of optically thick PSCs, especially over the tropics, then their greenhouse effect might have contributed to deglaciation, similar to the role of tropospheric clouds explored in \citet{abbot12-snowball-clouds,abbot2014}.

The changes in the simulated stratosphere of the snowball Earth are analogous but opposite to changes to the modern stratosphere that are predicted to occur in response to warming due to anthropogenic climate change. For example, the snowball involves weaker Brewer Dobson circulation, weaker eddy fluxes, zero SSWs, a stronger stratospheric vortex (under modern ozone), and colder temperatures compared to modern. In contrast, during warmer climates, the circulation is predicted to accelerate due to increased eddy fluxes \citep{mclandress2009simulated}. {In a warmer climate there are competing effects of a colder stratosphere and more wave driving \citep{ayarzaguena2018no}; however, in a model with a strengthened Madden Julian Oscillation, the stratosphere is predicted to exhibit a weaker polar vortex, more sudden stratospheric warming events, and warmer temperatures \citep{kang2017more}}.

Given the bi-directional coupling between stratospheric and tropospheric dynamics that has been observed and modeled under modern conditions, an altered snowball stratosphere may impact deglaciation scenarios and geochemical proxy interpretation. Interesting directions for future investigation include the modeling of stratospheric cloud formation and the inclusion of dynamical ozone chemistry in 3D simulations of the snowball atmosphere. It would also be useful to examine the stratospheric properties of partially glaciated climates like the Jormungand \citep{Abbot-et-al-2011:Jormungand}, the Slushball \citep{Micheels08}, or a stable ``Waterbelt'' \citep{rose2015stable}, which might have significantly different dynamics and trace chemical abundances, with the potential to influence interpretation of the $\Delta^{17}$O CO$_2$ proxy.

\begin{acknowledgments}
 This work was completed using resources provided by the University of Chicago Research Computing Center. We thank two anonymous reviewers and Dann Mitchell for their insightful comments. We acknowledge support from the University of Chicago College Research Fellows Program, NASA grant number NNX16AR85G, which is part of the {``}Habitable Worlds{''} program, the Packard Foundation, the Clarendon Scholarship, and Modalert. Data to reproduce figures can be found at DOI:10.6082/uchicago.1920.
\end{acknowledgments}

\bibliographystyle{agufull08}
\bibliography{thebibliography}


\clearpage
\begin{table}

\centering

\makebox[\textwidth][c]{\begin{tabular}{c r c c|r r r r r}
\hline
 Climate & CO$_2$ (ppmv)  & Topography &  Ozone& $\Psi_{\text{STEM}}^{\text{tropo}}$[$10^{9}\frac{\textrm{kg}}{\textrm{s}}$] & $\Psi_{\text{STEM}}^{\text{strat}}$[$10^{9}\frac{\textrm{kg}}{\textrm{s}}$]& $\gamma$ [$yr^{-1}$] &$\Theta$&$\gamma\times\Theta$[$yr^{-1}$]\\
\hline 
  Modern & 280  & Modern & Modern &269& 12.6&0.12& 0.018&0.0022\\
  Snowball & 280  & Modern &Modern &541&8.3 & {0.10}&0.026&{0.0026}\\
  Snowball & 10,000  &Modern&  Modern&631&10.3 &{0.11} &0.022&{0.0024}\\
  Snowball & 280 & Modern&  None&569 & 7.8& {0.082}&0.037&{0.0030}\\
  Snowball & 280 & Flat&  Modern&550& 6.9& {0.099}& 0.025&{0.0025}\\
\hline
\end{tabular}}
\caption{A table presenting information about the simulations used in this study. The ``Climate'' column classifies simulations as snowball or modern, ``CO$_2$ (ppmv)'' quantifies the CO$_2$ concentration of each simulation, ``Topography'' distinguishes between the flattened simulation and those with modern topography, ``Ozone'' describes the ozone content of each simulation, ``$\Psi_{\text{STEM}}^{\text{tropo}}$[$10^{9}\frac{\textrm{kg}}{\textrm{s}}$]'' gives the maximum absolute value reached by each STEM circulation averaged over December, January, and February,``$\Psi_{\text{STEM}}^{\text{strat}}$[$10^{9}\frac{\textrm{kg}}{\textrm{s}}$]'' gives the maximum absolute value reached by each STEM circulation above the isentrope marking the stratospheric ``overworld'' (380 K for modern, 285 K for snowballs) averaged over December, January, and February, ``$\gamma$ [$yr^{-1}$]'' provides the cross-tropopause mass exchange rates we calculated, ``$\Theta$'' provides the stratospheric mixing efficiencies, and ``$\gamma\times\theta$[$yr^{-1}$]'' lists the products of $\gamma$ and $\Theta$ in each simulation. 
\label{tab:sims}
}
\end{table}

\begin{table}
\centering
\resizebox{\textwidth}{!}
{
\begin{tabular}
{
c r c c |r r r r r r|r r r r r r
}
\hline
\multicolumn{16}{c}{\textbf{Winter cell}}\\
\hline
\text{Climate} & CO$_2$ &Topog. & Ozone &\multicolumn{6}{c|}{\text{Troposphere}[$10^{9}\frac{\textrm{kg}}{\textrm{s}}$]}&\multicolumn{6}{c}{\text{Stratosphere}[$10^{9}\frac{\textrm{kg}}{\textrm{s}}$]}\\
& (ppmv)&&&$\Psi_{\text{STEM}}$ & $\Psi_{\text{eddy}}$ & $\Psi_{\text{mean}}$ & $\Psi_{\text{adv}}$ & $\Psi_{\text{EMFC}}$ & $\Psi_{\text{{F}}}$ & $\Psi_{\text{STEM}}$ & $\Psi_{\text{eddy}}$ & $\Psi_{\text{mean}}$ & $\Psi_{\text{adv}}$& $\Psi_{\text{EMFC}}$ & $\Psi_{\text{{F}}}$\\
\hline
Modern & 280  & Modern & Modern &  -246.9&3.0&-259.9&-83.3&-158.1&-8.5&-12.6&-1.6&-11.0&-1.6&-1.3&-8.0\\
  Snowball & 280  & Modern &Modern&-428.6 &-8.7&-420.0&-186.7&-101.1&-132.1&-8.3&-0.5&-7.8&-1.0&-0.2&-6.6\\
  Snowball & 10,000  &Modern&  Modern& -468.0&-8.0&-460.0&-236.4&-107.4&-116.2&-10.3&0.4&-10.7&-3.1&-5.0&-2.6\\
  Snowball & 280 & Modern&  None&-431.1&-11.0&-420.1&-152.7&-126.9&-140.6&-7.8&-5.4&-2.4&-1.6&4.4&-5.2\\
  Snowball & 280 & Flat&  Modern&-454.4&-7.0&-447.5&-201.8&-116.1&-129.6&-6.9&-2.4&-4.5&-1.4&-1.6&-1.5\\
\hline
\end{tabular}
}
\caption{The strength of the DJF-mean STEM circulation in the stratosphere and the troposphere of the winter hemisphere is decomposed. The maximum strength of the STEM circulation outside of latitudes between 8.4 and $-8.4^\circ$ (where $\Psi_\text{adv}$, $\Psi_\text{EMFC}$, and $\Psi_\text{{F}}$ are whited out in Figs. {\ref{fig:era}} through \ref{fig:modern} and \ref{fig:280ppmv} through \ref{fig:nomount_difference}) for an atmospheric layer is listed under each $\Psi_\text{STEM}$ subcolumn. We take the bottom boundary for inclusion in the stratosphere to be 285 K for snowball simulations and 380 K for the modern simulation. $\Psi_{\text{eddy}}$ through $\Psi_{\text{{F}}}$ lists the strength of each component at the same latitude and potential temperature as the $\Psi_\text{STEM}$ value. Note that these values are $\textit{not}$ the extrema of $\Psi_\text{eddy}$ through $\Psi_\text{{F}}$: they are the values of those circulation components at the extremum of $\Psi_\text{STEM}$. The $\Psi_\text{STEM}$ values in the ``Troposphere'' column do not match those of Table \ref{tab:sims} because the actual maximum of the Hadley circulation often occurs within +/-8.4 $^\circ$. Columns ``Climate'' through ``Ozone'' are the same as in Table \ref{tab:sims}. 
\label{tab:psi_comps_winter}
}
\end{table}

\begin{table}
\centering
\resizebox{\textwidth}{!}
{
\begin{tabular}
{
c r c c |r r r r r r|r r r r r r
}
\hline
\multicolumn{16}{c}{\textbf{Summer cell}}\\
\hline
\text{Climate} & CO$_2$ &Topog. & Ozone &\multicolumn{6}{c|}{\text{Troposphere}[$10^{9}\frac{\textrm{kg}}{\textrm{s}}$]}&\multicolumn{6}{c}{\text{Stratosphere}[$10^{9}\frac{\textrm{kg}}{\textrm{s}}$]}\\
& (ppmv)&&&$\Psi_{\text{STEM}}$ & $\Psi_{\text{eddy}}$ & $\Psi_{\text{mean}}$ & $\Psi_{\text{adv}}$ & $\Psi_{\text{EMFC}}$ & $\Psi_{\text{{F}}}$ & $\Psi_{\text{STEM}}$ & $\Psi_{\text{eddy}}$ & $\Psi_{\text{mean}}$ & $\Psi_{\text{adv}}$& $\Psi_{\text{EMFC}}$ & $\Psi_{\text{{F}}}$\\
\hline
Modern & 280  & Modern & Modern &105.8 &14.0&91.7&24.5&67.2&0.0&6.8&7.7&-0.9&0.3&-3.0&1.7\\
  Snowball & 280  & Modern &Modern& 12.8&12.0 &0.8&-0.1&-0.8&1.6&2.2&2.1&0.1&-0.2&0.2&0.1\\
  Snowball & 10,000  &Modern&  Modern&9.5&9.8&-0.2&-0.5&-0.1&0.4&2.2&2.0&0.2&-0.3&0.1&0.3\\
  Snowball & 280 & Modern&  None&17.4&16.4&1.0&0.0&0.1&0.9&1.3&1.2&0.2&0.0& 0.0&0.1\\
  Snowball & 280 & Flat&  Modern&30.8&29.5&1.2&0.0&0.6&0.7&2.9&2.6&0.3&-0.2&0.3&0.2\\
\hline
\end{tabular}
}
\caption{Same as Table \ref{tab:psi_comps_winter}, except presenting information from the summer hemisphere of DJF-averaged simulations.
\label{tab:psi_comps_summer}
}
\end{table}

\end{document}